\documentclass[aps,11pt,nofootinbib]{revtex4-1}
\pdfoutput=1
\usepackage{graphicx}
\usepackage[margin=1in]{geometry}
\usepackage{amsmath}
\usepackage{amssymb}
\usepackage{color}
\usepackage[utf8]{inputenc}
\usepackage{hyperref}
\usepackage[normalem]{ulem} 
\usepackage{cancel} 
\usepackage{comment}

\makeatletter
\renewcommand\@make@capt@title[2]{%
  \@ifx@empty\float@link{\@firstofone}{\expandafter\href\expandafter{\float@link}}%
   {\textbf{#1}}\@caption@fignum@sep#2\quad
}%
\makeatother

\newcommand{\be}{\begin{equation}}
\newcommand{\ee}{\end{equation}}

\begin{document}

\title{
Computing the abundance of primordial black holes
}

\author{Sam Young}
\email{young@lorentz.leidenuniv.nl}
\affiliation{Instituut-Lorentz for Theoretical Physics, Leiden University,\\Niels Bohrweg 2, 2333 CA Leiden, The Netherlands}


\begin{abstract}
An accurate calculation of their abundance is crucial for numerous aspects of cosmology related to primordial black holes (PBHs). For example, placing constraints on the primordial power spectrum from constraints on the abundance of PBHs (or vice-versa), calculating the mass function observable today, or predicting the merger rate of (primordial) black holes observable by gravitational wave observatories such as LIGO, Virgo and KAGRA.
In this chapter, we will discuss the different methods used for the calculation of the abundance of PBHs forming from large-amplitude cosmological perturbations, assuming only a minimal understanding of modern cosmology. Different parameters to describe cosmological perturbations will be considered (including different choices for the window function), and it will be argued that the compaction is typically the most appropriate choice. Different methodologies for calculating the abundance and mass function are explained, including \emph{Press-Schechter}-type and peaks theory approaches.
\end{abstract}

\maketitle

{
  \hypersetup{linkcolor=black}
  \tableofcontents
}

\normalem

\section{Introduction}
\label{sec7:intro}

Primordial black holes (PBHs) have a wealth of potential implications for cosmology - from placing unique constraints on the primordial power spectrum, solving numerous unexplained phenomena (such as the early formation of supermassive black holes (SMBHs)), and the observed presence of dark matter in the Universe. Drawing robust conclusions in such scenarios typically relies on a calculation of quantities such as the abundance of PBHs from given initial conditions, the masses at which they form, and their initial clustering. Making this calculation as accurate and reliable as possible is therefore crucial.

\begin{figure}
\centering
\includegraphics[width=0.7\columnwidth]{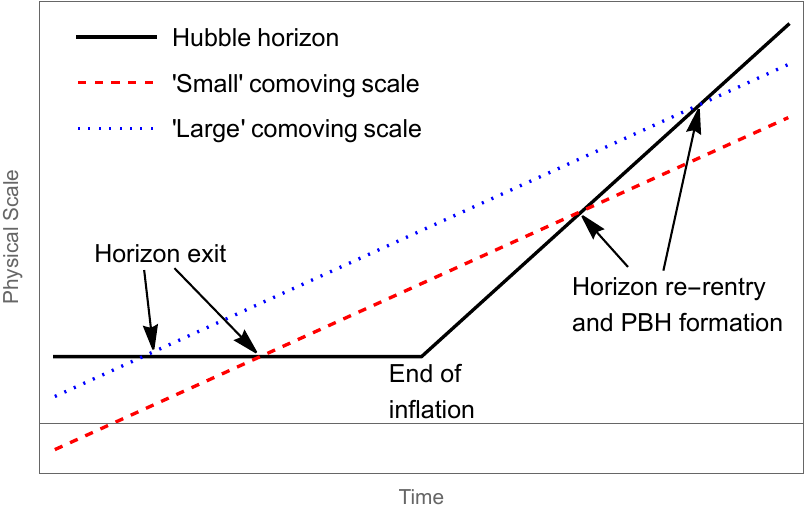}
\caption{\emph{Horizon exit and re-entry.} The black line shows the Hubble horizon as a function of time, remaining constant during (quasi-)exponential inflation, before growing during the radiation dominated epoch following reheating. The dashed red and dotted blue lines show how perturbations of a small and large scale, respectively, first exit the horizon during inflation before re-entering at a later time. PBHs can form upon horizon re-entry if the amplitude of a perturbation is large enough.}
\label{fig7:horizons}   
\end{figure}

In this chapter, we will discuss the calculation of the abundance of PBHs from the collapse of large density fluctuations during radiation domination, but let us first turn our attention to the cosmological setting in which such PBHs form. During (exponential) inflation, the physical Hubble horizon is constant, and quantum fluctuations become classicial density perturbations as they exit the horizon as the universe expands. Once inflation ends and the universe becomes radiation dominated, the Hubble horizon starts to grow, and modes which had previously exited the horizon re-enter.
Figure \ref{fig7:horizons} shows a schematic plot of how perturbations exit and re-enter the horizon. Larger-scale perturbations exit the horizon earlier, and then re-enter the horizon later.
Once a perturbation re-enters the horizon, if the amplitude of the perturbation is large enough, gravity can overcome pressure forces, causing it to quickly collapse and form a PBH~\cite{Zeldovich:1967,Hawking:1971ei,Carr:1975qj}, with a mass of approximately the horizon mass, or else be quickly damped out if the perturbation is not dense enough. This means that larger-scale perturbations correspond to earlier times during inflation, and more massive PBHs which form at later times.

\begin{figure}
\centering
\includegraphics[width=0.7\columnwidth]{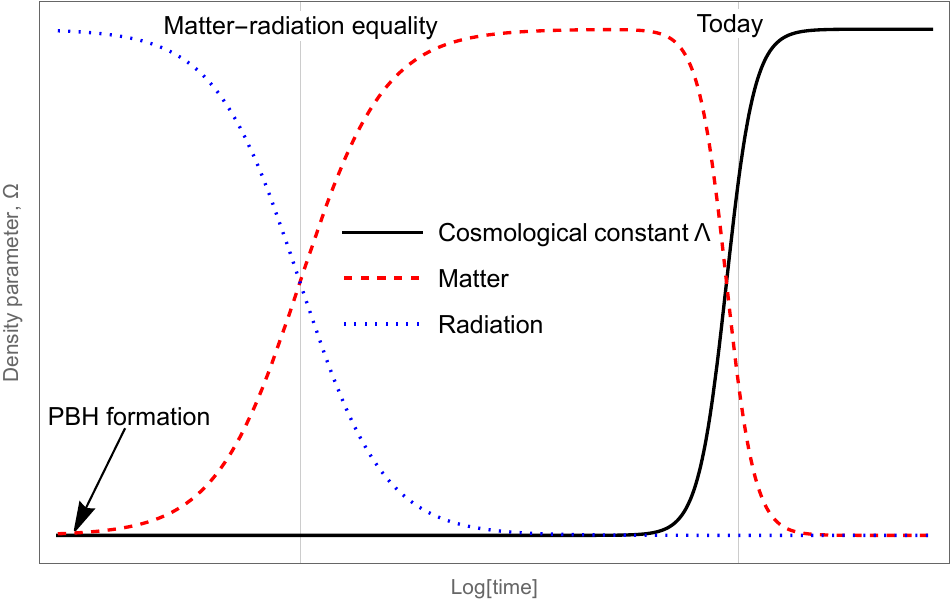}
\caption{\emph{The evolution of cosmological densities.} The figure show the relative energy density of different species within the universe. Initialy, following reheating, the energy density of the universe was dominated by radiation. As the universe expands, the radiation density decreases faster than the matter density, and the universe enter a matter dominated epoch. Finally, the universe becomes dominated by the cosmological constant, whose energy density remains constant. Because PBHs act like matter, and form very early during radiation domination, the fraction of the  energy of the universe contained in PBHs is typically expected to be very small (expect in models where such PBHs evaporate extremely rapidly).}
\label{fig7:density}   
\end{figure}

Once an abundance of PBHs has formed, they behave on large scales like a pressure-less fluid, and the PBH density grows inversely proportional to the cube of the scale factor,
\begin{equation}
    \rho_\mathrm{PBH} \propto a^{-3},
\end{equation}
i.e. the PBH density on large scales evolves like matter,
whilst the density of the radiation fluid grows inversely to the fourth power of the scale factor,
\begin{equation}
    \rho_\mathrm{r}\propto a^{-4}.
\end{equation}
Although not relevant for our discussion here, the final quantity which makes up a significant fraction of the universe today is dark energy, which we will assume is due to a cosmological constant $\Lambda$. The density of the cosmological constant remains constant over time,
\begin{equation}
    \rho_\Lambda\propto \mathrm{const.}
\end{equation}
The density parameter $\Omega_i$ describes the density of a particular fluid $\rho_i$ relative to the total density $\rho_{T}$,
\begin{equation}
    \Omega_i = \frac{\rho_i}{\rho_T}.
\end{equation}
Figure \ref{fig7:density} shows how the density parameter changes over time for matter, radiation and the cosmological constant. PBHs form very early during radiation domination, at a time when the (relative) density of matter is very low. Because the abundance of PBHs cannot be larger than the abundance of matter, with the exception of very light PBHs which evaporate very quickly, the abundance of PBHs in the universe at the time of formation is typically assumed to be very small: PBHs are extremely rare at the time of formation.

For PBHs to form, significantly larger perturbations are needed than are observed on large scales in e.g. the cosmic microwave background radiation (CMB). Perturbations are required to be approximately $50\%$ more dense than the background~\cite{Musco:2004ak,Musco:2008hv,Nakama:2014bxa,Nakama:2014xwa}. A consequence of this is that the primordial power spectrum must be orders of magnitude larger on small scales than the value of $\sim 2\times 10^{-9}$ observed in the CMB~\cite{Planck:2018jri}, with the typical values quoted in the literature around $\mathcal{O}(10^{-2})$ (see e.g.~\cite{Carr:2020gox}). However, there can be significant deviation in this value depending on the method used (e.g.\cite{Gow:2020bzo}).

In this chapter, we will first discuss how cosmological perturbations can be described and how they evolve over time. We will then discuss the typical profile shapes of perturbations, and how this can affect the calculation of the PBH abundance. We will then explain the different methods which can be used to calculate the PBH abundance, beginning with a simple \emph{Press-Schechter}-type approach, before moving on to the theory of peaks and its extensions. Finally, we will consider the effect of non-linearities in the density, which become important as large perturbations are required for PBH formation.

\section{Describing cosmological perturbations}
\label{sec7:perturbations}

Many models of inflation can predict a large number of PBHs to form in the early universe, including the running mass model \cite{Drees:2011hb}, axion inflation \cite{Bugaev:2013fya,Ozsoy:2018flq}, or a waterfall transition during hybrid 
inflation \cite{GarciaBellido:1996qt,Lyth:2012yp,Bugaev:2011wy}, and many others. 
The curvature perturbation $\zeta$ is typically used to study cosmological perturbations generated with the different models and to predict their observable consequences (NB. Often
 $\mathcal{R}$ is used instead to denote the curvature perturbation, whilst $\zeta$ is used to describe the curvature perturbation only on a uniform density slicing), and is also what is typically constrained by large-scale observations, such as the \emph{Planck} satellite observations of the CMB~\cite{Planck:2018jri}. The curvature perturbation $\zeta$ appears as a perturbative quantity in the Friedmann-Lemaitre-Robertson-Walker (FLRW) metric in the comoving uniform-density gauge as
\begin{equation} 
\label{eqn7:FLRWmetric}
\mathrm{d} s^2 =-\mathrm{d}t^2 + a^2(t)\exp(2\zeta)\mathrm{d}\mathbf{X}^2,
\end{equation}
where $t$ is time, $a(t)$ is the scale factor and $\mathbf{X}$ represents the three comoving spatial coordinates. To use constraints on PBHs to constrain models of inflation (or to make predictions of PBH abundances from a given model), we therefore need to relate the primordial curvature perturbation power spectrum $\mathcal{P}_\zeta(k)$ to the abundance of PBHs of different masses.

However, the curvature perturbation itself is not a suitable parameter to determine PBH formation. Whilst it is possible to state a threshold value for $\zeta$ above which a fluctuation will collapse to form a PBH~\cite{Shibata:1999zs}, this assumes that the curvature perturbation goes to zero outside the scale of the perturbation. This is not generally the case: small-scale perturbations will typically be superposed on top of large-scale perturbations. Naively using the curvature perturbation as the formation threshold would therefore suggest that larger-scale perturbations have a significant impact on PBH formation. However, precisely because such perturbations are well outside the horizon at the time of PBH formation, the separate universe approach suggests that it should not affect the local evolution of the universe~\cite{Wands:2000dp}. This can be seen by separating the curvature perturbation into a small horizon-scale perturbation $\zeta_s$ and a large super-horizon perturbation $\zeta_l$, and then treating the large-scale super-horizon mode as a constant:
\begin{equation}
    \zeta=\zeta_s+\zeta_l.
\end{equation}
We then consider the term in the FLRW metric where the curvature perturbation appears,
\begin{equation}
    a(t)\exp(\zeta)\rightarrow a(t)\exp(\zeta_s+\zeta_l) = \left[ a(t)\exp(\zeta_l) \right]\exp(\zeta_s).
\end{equation}
If we then consider that $\zeta_l$ is (very close to) constant on the horizon scale, we can treat that term as simply a rescaling of the scale factor, corresponding to a time shift $t'$ in the background evolution of the universe,
\begin{equation}
    a(t)\rightarrow a(t)\exp(\zeta_l) = a(t+t').
\end{equation}
We conclude that, super-horizon modes in $\zeta$ should not affect whether a PBH forms or not, but can have an effect at the time at which a PBH would form.

Seemingly the most obvious parameter to use is then the density contrast $\delta$, defined as 
\begin{equation}
    \delta(t,\mathbf{x})\equiv \frac{\rho-\bar\rho}{\bar\rho} 
    \label{eqn7:denCon}
\end{equation}
which, at first order\footnote{We will consider the effect of non-linearities arising from higher order terms in section \ref{sec7:non-linearities}.}, is related to $\zeta$ as
\begin{equation}
    \delta(t,\mathbf{x}) = -\frac{2(1+\omega)}{5+3\omega}\left(\frac{1}{aH}\right)^2\nabla^2\zeta(\mathbf{x}),
\end{equation}
where $\mathbf{x}$ is the spatial coordinate, $\rho$ is the density, $\bar\rho$ is its background value, $\omega$ is the equation of state (we will consider that $\omega=1/3$ during radiation domination for the rest of this chapter), $H(t)=\dot{a}(t)/a(t)$ is the Hubble parameter, and $\nabla^2=\frac{\partial}{\partial x} +\frac{\partial}{\partial y}+\frac{\partial}{\partial z}$ is the Laplacian. The comoving Hubble horizon is given by $R_\mathrm{H}=(aH)^{-1}$. In Fourier space, this relation is given by
\begin{equation}
    \delta(t,k) = \frac{2(1+\omega)}{5+3\omega}\left(\frac{k}{aH}\right)^2\zeta(t,k).
    \label{eqn7:deltaLinear}
\end{equation}
On super-horizon scales, $\zeta$ is constant over time, $\zeta(t,k)\rightarrow \zeta(k)$, and so the term $\left(\frac{k}{aH}\right)^2$ contains the time-dependance of the density contrast: perturbations becoming larger over time as the horizon grows, and larger-scale modes are suppressed by a factor $k^2$. 

On sub-horizon scales, it is necessary to account for the evolution of perturbations, which, at linear order, is easily accounted for in Fourier space by including a transfer function $T(R_\mathrm{H},k)$ which includes the time-dependence,
\begin{equation}
    \delta(R_\mathrm{H},k) = \frac{2(1+\omega)}{5+3\omega}\left(\frac{k}{aH}\right)^2\zeta(k)T(R_\mathrm{H},k),
\end{equation}
with the transfer function given by~\cite{Josan:2009qn},
\begin{equation}
    T(R_\mathrm{H},k) = 3\frac{\sin\left(\frac{k R_\mathrm{H}}{\sqrt{3}}\right)-\frac{k R_\mathrm{H}}{\sqrt{3}}\cos\left(\frac{k R_\mathrm{H}}{\sqrt{3}}\right)}{\left(\frac{k R_\mathrm{H}}{\sqrt{3}}\right)^3},
\end{equation}
where the Hubble horizon scale $R_\mathrm{H}(t)$ is a function of time. However, it will be more convenient for our purposes to simply use $R_\mathrm{H}$ itself as the time parameter. On super-horizon scales, $k\ll R_\mathrm{H}$, and the transfer function approach unity, recovering equation \eqref{eqn7:deltaLinear}, and is highly oscillatory on sub-horizon scales.

It is common to use the density contrast as the formation criterion: for any perturbations where the density peaks above some threshold $\delta_{c}$, it is considered that a PBH will form. However, simulations reveal that the formation (or not) of a PBH is best determined by the average density within the horizon, rather than the peak value (likely at the center of the perturbation)~\cite{Musco:2018rwt}, which brings us to the discussion of smoothing functions (also referred to as window functions, or simply windows).



\subsection{Smoothing functions}
\label{sec7:smoothing}

A smoothing function can be considered as an averaging function - it returns the average value of a quantity within a specified volume, centred on a certain coordinate. In order to calculate the smoothed density contrast $\delta_R$, a smoothing function $W(R_s,\mathbf{x})$ is convolved with the density contrast:
\begin{equation}
    \delta_R(t,\mathbf{x},R_s) = \int\mathrm{d}^3\mathbf{y} \delta(t,\mathbf{x}-\mathbf{y})W(R_s,\mathbf{y}),
    \label{eqn7:smoothing}
\end{equation}
where $R_s$ is the smoothing scale, and the integration is performed over all space (or at least, the full extent of the smoothing function). In Fourier space, the density is expressed simply as a multiplication with the smoothing function:
\begin{equation}
    \delta_R(t,k) = \delta(t,k)\tilde{W}(k,R),
    \label{eqn7:fourierSmoothing}
\end{equation}
where the tilde denotes a Fourier transform.

A smoothing function will typically define a volume of space to be integrated over (potentially giving different weight to different regions), divided by the volume. Conceptually, the simplest smoothing function is the top-hat window function $T_\mathrm{TH}$,
\begin{equation}
    W_\mathrm{TH}(\mathbf{x},R_s) = \frac{3}{4\pi R_s^3}\Theta_H(R-x),
    \label{eqn7:topHat}
\end{equation}
where $\Theta_H(x)$ is the Heaviside theta function, and $x=\left|\mathbf{x}\right|$.
We can insert this into equation \eqref{eqn7:smoothing} to obtain
\begin{equation}
    \delta_R(t,\mathbf{x},R_s) = \frac{3}{4\pi R_s^3}\int\mathrm{d}^3\mathbf{y} \delta(t,\mathbf{x}-\mathbf{y})\Theta_H(R-y).
\end{equation}
We further simplify by choosing to center our coordinates on $\mathbf{x}$ and assuming spherical symmetry around this point,
\begin{equation}
    \delta_R(t,0,R_s) = \frac{3}{4\pi R_s^3}\int\limits_0^{R_s}\mathrm{d}\mathbf{r} 4\pi r^2 \delta(t,r),
\end{equation}
where we have used $\mathrm{d}^3y = 4\pi r^2\mathrm{d}r$, which is valid under spherical symmetry. It is clear, here, that the smoothed value $\delta_R$ is simply $\delta$ integrated over a sphere of radius $R_s$, divided by the volume of the sphere i.e. the volume-averaged value.

In addition to the top-hat smoothing function, another common choice is the Gaussian smoothing function,
\begin{equation}
    W_G(\mathbf{x},R_s) = \frac{1}{\left(\pi R_s^2\right)^{3/2}}\exp\left( \frac{-x^2}{R_s^2} \right).
\end{equation}
We note that, often, there is an additional factor of $1/2$ included in the exponent of this function which has been neglected here. This is a free choice, but is made here such that the characteristic scales of the Gaussian and top-hat window functions coincide (i.e. the integrand $4\pi r^2W(r,R_s)$ peaks at $R_s$ in both cases).

\begin{figure}
\centering
\includegraphics[width=0.95\columnwidth]{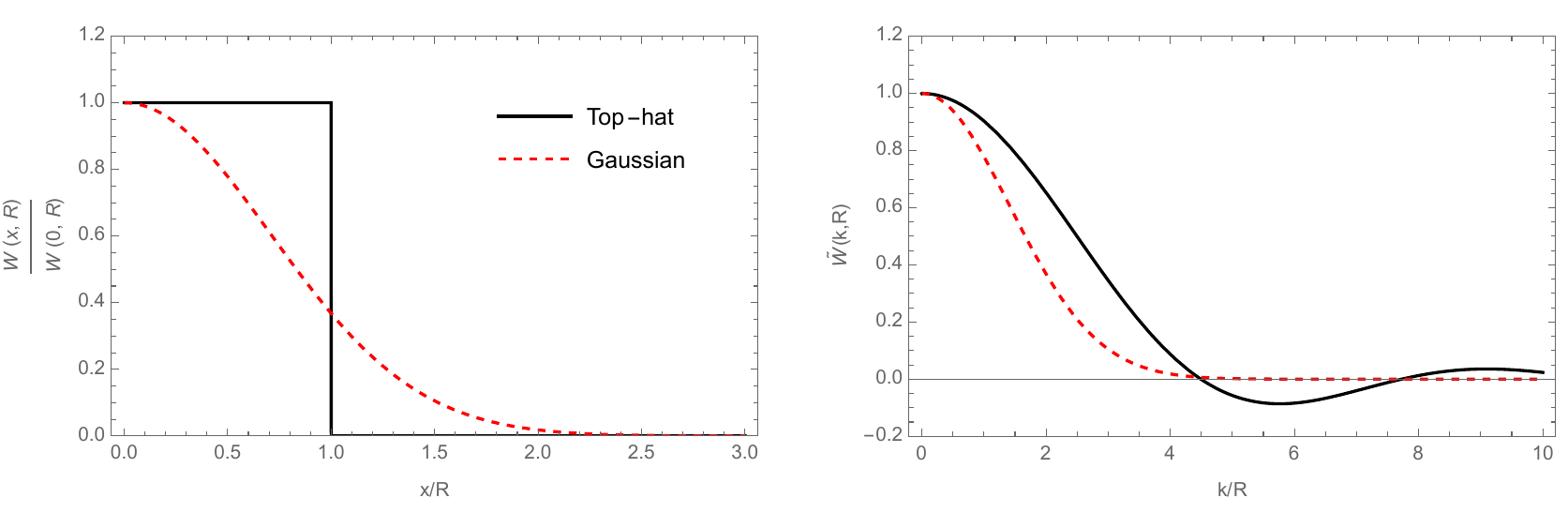}
\caption{\emph{Smoothing functions.} The shape of top-hat and Gaussian smoothing functions are shown in real-space (left) and Fourier-space (right). }
\label{fig7:smoothing}   
\end{figure}

The Fourier transforms of these smoothing functions are given by
\begin{equation}
    \tilde{W}_\mathrm{TH}(k,R_s) =  3\frac{\sin\left(k R_\mathrm{H}\right)-\frac{k R_\mathrm{H}}{\sqrt{3}}\cos\left(\frac{k R_\mathrm{H}}{\sqrt{3}}\right)}{\left(\frac{k R_\mathrm{H}}{\sqrt{3}}\right)^3},
    \label{eqn7:fourierTH}
\end{equation}
\begin{equation}
    \tilde{W}_\mathrm{G}(k,R_s) = \exp\left( \frac{-k^2R_s^2}{4} \right),
    \label{eqn7:fourierGaussian}
\end{equation}
Figure \ref{fig7:smoothing} shows the real- and Fourier-space transforms of the top-hat and Gaussian smoothing functions. Which smoothing function to use is the subject of ongoing debate, but \emph{a priori} there is currently no reason to choose one smoothing over another - although both presented here have their own advantages and disadvantages which will be discussed later.

The use of a smoothing function has several benefits compared to using the unsmoothed density contrast. As previously mentioned, the threshold for collapse is more accurately stated in terms of the smoothed density contrast. Whilst the exact value depends on the shape of a the specific perturbation being considered, the critical value varies in the range $0.41\lesssim \delta_{R,c}\lesssim 0.67$ for extreme shapes of perturbations. However, the critical value for the peak value of the (unsmoothed) density contrast can, in principle vary between positive and negative infinity (e.g. for perturbations with a very sharp peak or trough at the centre)~\cite{Musco:2018rwt}. In addition, smoothing the distribution on different scales means that the abundance of PBHs forming on different scales (and, therefore, of different masses) can be calculated~\cite{Young:2019osy}.

\subsection{The compaction function}
\label{sec7:compaction}

Very closely related to the smoothed density contrast, the compaction function $C$ has, in recent years, become the standard parameter used to describe perturbations, and is defined as
\begin{equation}
    C(r,t,\mathbf{x}) = \frac{2\delta M(r,t,\mathbf{x})}{r},
\end{equation}
where $\delta M(r,\mathbf{x}) = M(r,t,\mathbf{x})-\bar{M}(r,t)$ is the mass excess within a sphere of areal radius $r$ centred on spatial coordinate $\mathbf{x}$; $M(r,t,\mathrm{x})$ is the Misner-Sharp mass and the bar denotes the background value in a region of unperturbed space. As we will see soon, the compaction is constant on superhorizon scales, and the time dependance is typically dropped. The mass excess $\delta M$ can be expressed as an integral over the density
\begin{equation}
    \delta M(r,t,\mathbf{x}) = \int \mathrm{d}^3\mathbf{y}\left[ \rho(t,\mathbf{x}-\mathbf{y}) -\bar{\rho}(t)\right]\Theta_H(r-y).
    \label{eqn7:massExcess}
\end{equation}
Using natural units, $c=G=1$, and assuming a flat universe, the Friedmann equation gives an expression for the background density,
\begin{equation}
    \bar{\rho}(t) = \frac{3 H^2(t)}{8 \pi}.
\end{equation}
Using the definition for the density contrast, equation \eqref{eqn7:denCon}, allows us to rewrite equation \eqref{eqn7:massExcess} as
\begin{equation}
    \delta M(r,t,\mathbf{x}) = \frac{3H^2}{4\pi}\int\mathrm{d}^3\mathbf{y}\delta(t,\mathbf{x}-\mathbf{y})\Theta_H(r-y).
\end{equation}
The compaction function can then be expressed as,
\begin{equation}
    C(r,t,\mathbf{x}) = \frac{r^2}{r_\mathrm{H}^2} \int\mathrm{d}^3\mathbf{y}\delta(t,\mathbf{x}-\mathbf{y})\frac{3}{4\pi r^3}\Theta_H(r-y),
\end{equation}
where $r_\mathrm{H}=H^{-1}$ is the physical Hubble radius. The term inside the integral is the top-hat smoothing function, equation \eqref{eqn7:topHat}, so finally we can write
\begin{equation}
    C_\mathrm{TH}(r,t,\mathbf{x}) = \frac{r^2}{R_\mathrm{H}^2}\int\mathrm{d}^3\mathbf{y}\delta(t,\mathbf{x}-\mathbf{y})W_\mathrm{TH}(r,\mathbf{y}),
    \label{eqn7:Cth}
\end{equation}
which is exactly the same as the smoothed density contrast, equation \eqref{eqn7:smoothing}, where $r$ has taken the role of the smoothing scale, and multiplied by $r^2/r_\mathrm{H}^2$. The compaction function can also be defined using a Gaussian smoothing:
\begin{equation}
    C_\mathrm{G}(r,t,\mathbf{x}) = \frac{r^2}{R_\mathrm{H}^2}\int\mathrm{d}^3\mathbf{y}\delta(t,\mathbf{x}-\mathbf{y})W_\mathrm{G}(r,\mathbf{y}).
\end{equation}

The compaction function is therefore extremely closely related to the smoothed density contrast, and indeed, the two are often used interchangeably in the literature (see e.g. reference~\cite{Young:2019osy}). This is especially true as the smoothed density contrast is often evaluated at horizon entry, such that $r/r_\mathrm{H}=1$. However, the compaction is typically preferred as it is conserved on super-horizon scales. This can be seen by inserting equation \eqref{eqn7:denCon}, and replacing the physical radius $r=aR$ with the comoving radius $R$:
\begin{equation}
    C(R,\mathbf{x}) = -R^2\frac{2(1+\omega)}{5+3\omega}\int\mathrm{d}^3\mathbf{y}\nabla^2\zeta(\mathbf{x}-\mathbf{y})W(R,\mathbf{y}),
    \label{eqn7:compationZeta}
\end{equation}
where the time-dependance has been dropped, because $\zeta$ is conserved on super-horizon scales. 

It is typically simpler to leave this integral as it appears, as the expressions in Fourier space are quite simple (e.g. equation \eqref{eqn7:fourierSmoothing}). However, it will be useful later if we perform the integration analytically for the case of the top-hat smoothing, which can be completed under the assumption of spherical symmetry. This assumption is typically justified because, as discussed in section \ref{sec7:intro}, PBHs are extremely rare at the time of formation - which therefore means that the perturbations from which they form are very rare, due to the large amplitude required. The theory of peaks tells us that, the larger a perturbation is, the more spherical symmetry it will have~\cite{Bardeen:1985tr}\footnote{However, we note that it has been pointed out that, whilst the perturbations to the density (or compaction) are large and rare, this does not necessarily correspond to large, rare perturbations in $\zeta$~\cite{Young:2022phe}.}. Under the assumption of spherical symmetry, the Laplacian $\nabla^2\zeta$ becomes
\begin{equation}
    \nabla^2\zeta = \zeta''(r)+\frac{2}{r}\zeta'(r),
\end{equation}
where $r$ is here the radial coordinate, and the prime denotes a derivative with respect to $r$. The compaction can then be expressed as
\begin{equation}
    C(R) = -\frac{2(1+\omega)}{5+3\omega}\frac{3}{R}\int\limits_0^R \mathrm{d}r r^2\left[ \zeta''(r)+\frac{2}{r}\zeta'(r) \right] = -\frac{6(1+\omega)}{5+3\omega}R\zeta'(R).
    \label{eqn7:C1zeta}
\end{equation}
To avoid confusion, we stress here that, the $\zeta'$ term has arisen as the surface term of an integral over a spherical volume - therefore corresponding to $\zeta'$ smoothed over a spherical shell.

\section{Perturbation profiles}
\label{sec7:profiles}

The single most important parameter when considering the abundance of PBHs which will form from the collapse of cosmological perturbations is the threshold amplitude of the perturbations required for collapse, $C_c$. Carr's initial order-of-magnitude estimate, calculated using a Jeans' Length argument, was that the threshold for collapse is equal to the equation of state, $C_c\sim\omega$~\cite{Carr:1974nx,Carr:1975qj}, which is equal to $1/3$ during radiation domination. Since then, many more complex methods have been developed, utilising either analytic methods (e.g.~\cite{Harada:2015ewt,Escriva:2020tak}), or numerical simulations (e.g.~\cite{Musco:2004ak,Polnarev:2006aa,Musco:2008hv,Nakama:2013ica,Escriva:2019nsa,Escriva:2019phb}). The precise value depends on the initial profile shape of the perturbation which forms a PBH, which is a subject of immense scrutiny. In this section, we will briefly turn our attention towards the profile shape one might therefore expect for cosmological perturbations.

\begin{figure}
\centering
\includegraphics[width=0.7\columnwidth]{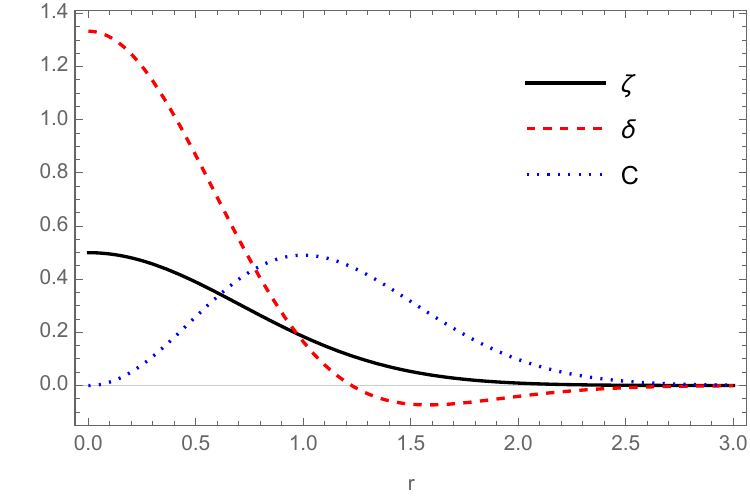}
\caption{\emph{Parameterising a perturbation.} The same perturbation is shown, described with several different parameters: the curvature perturbation $\zeta$, the density contrast $\delta$ and the compaction $C$. For $\zeta$ and $\delta$ the plot shows the value at the radial coordinate $r$. For the compaction, the plot shows the amplitude of the compaction at the centre of the perturbation, when smoothed with a smoothing scale $r$.}
\label{fig7:perturbation}   
\end{figure}

We will first consider how a single perturbation may be described using the various parameters discussed in section \ref{sec7:perturbations}, and we will consider only the initial perturbation on super-horizon scales, before any complicated evolution close to horizon entry. We therefore set the transfer function equal to unity, $T(t,k)=1$. We will take a spherically symmetric perturbation in $\zeta$ in an otherwise empty universe, given by a Gaussian profile:
\begin{equation}
    \zeta(r) = A \exp\left( -\frac{r^2}{r_p^2} \right),
    \label{eqn7:zetaProfile}
\end{equation}
where $A$ describes the amplitude of the perturbation, $r_p$ determines the scale of the perturbation, and $r$ is the comoving radial coordinate. The same perturbation can be described in terms of the density contrast $\delta$ and results in the ubiquitous Mexican-hat profile,
\begin{equation}
    \delta(r) = -\frac{4(1+\omega)}{5+3\omega}\left(\frac{1}{aH}\right)^2 A \frac{\exp\left( -\frac{r^2}{r_p^2} \right) \left( 2 r^2 - 3r_p^2 \right)}{r_p^4},
    \label{eqn7:mexicanHat}
\end{equation}
as well as the compaction $C$ (for a top-hat smoothing),
\begin{equation}
    C(r) = -\frac{12(1+\omega)}{5+3\omega} A\frac{r^2}{r_p^2}\exp\left( -\frac{r^2}{r_p^2} \right).
\end{equation}
Figure \ref{fig7:perturbation} shows the profile shapes in each of these parameters, where we have made the following choices, $\left[A,r_p,(aH)^{-1} \right]=\left[ 0.5,1,1\right]$. $A$ is chosen to correspond approximately to the amplitude for PBH formation, whilst the choice of $r_p$ is arbitrary, and $(aH)^{-1}$ is chosen so that the amplitude of the density is similar to the others (technically, in the super-horizon limit, we should have chosen $(aH)^{-1}\ll r_p$). For $\zeta$ and $\delta$, the profile simply represents the value of that parameter at the coordinate $r$, and this is also often considered the case for $C$. However, it is more accurate to consider the shown profile corresponds to the value of the compaction at the centre, when a smoothing scale $R_s=r$ has been used.

For any profile shape, the compaction is generically zero at $r\rightarrow 0$ and $r\rightarrow \infty$, and peaks at some value of $r$. The value at which this occurs is used to define the characteristic scale $r_m$ of the perturbation: $C'(r_m)\equiv 0$, and can be found by solving
\begin{equation}
    \zeta'(r_m)+r_m \zeta(r_m)=0.
\end{equation}
The characteristic scale $r_m$ will be important later as it corresponds to the scale (and mass) at which a PBH forms. In this case, with the profile given in equation \eqref{eqn7:zetaProfile}, we find that $r_m$ coincides with our choice of parameter for the scale, $r_m=r_p$.

We will now turn our attention to the consideration of what profile shapes we might expect to see in the density in different scenarios - as this will affect the amplitude a perturbation needs to have in order to form a PBH. The expected profile shape can be calculated by considering the amplitude of different Fourier modes in the universe~\cite{Bardeen:1985tr}, which is described by the power spectrum. The power spectrum $P_\zeta(k)$ is the Fourier transform of the 2-point correlation function,
\begin{equation}
    \langle \zeta(\mathbf{k}_1)\zeta(\mathbf{k}_2) \rangle = (2\pi)^3\delta_D^{(3)}(\mathbf{k}_1+\mathbf{k}_2)P_\zeta(k), 
\end{equation}
where $\delta_D^{(3)}$ is the (3D) Dirac-delta function. We will consider the dimensionless power spectrum $\mathcal{P}_\zeta(k)$, given by
\begin{equation}
    \mathcal{P}_\zeta(k) = \frac{k^3}{2\pi^2}P_\zeta(k).
\end{equation}
The effect of the power spectrum on perturbation profiles is considered in detail in the context of PBH formation by references~\cite{Germani:2018jgr,Young:2019osy}. In this case, we will consider density profiles, and so consider the density power spectrum, $\mathcal{P}_\delta(k)$, which is related to the curvature perturbation power spectrum, $\mathcal{P}_\zeta(k)$, as
\begin{equation}
    \mathcal{P}_\delta(R_\mathrm{H},k) = \frac{4(1+\omega)^2}{(5+3\omega)^2} \left( k R_\mathrm{H} \right)^4 \mathcal{P}_\zeta(k)T^2\left( R_\mathrm{H},k \right),
    \label{eqn7:powerSpectra}
\end{equation}
where, again, we use the comoving Hubble horizon $R_\mathrm{H}$ as the parameter for time, and will here consider the initial perturbations long before horizon entry, such that $k\ll R_\mathrm{H}$, and $T(R_\mathrm{H},k)\approx 1$.

Assuming spherical symmetry, the profile shape close to the centre of a perturbation can be stated in terms of the radial coordinate $r$,
\begin{equation}
    \delta(R_\mathrm{H},r) = \delta(R_\mathrm{H},0) \frac{\xi(R_\mathrm{H},r)}{\xi(R_\mathrm{H},0)},
    \label{eqn7:averageProfile}
\end{equation}
where $\delta(R_\mathrm{H},0)$ is the amplitude at the centre of the perturbation.The function $\xi(R_\mathrm{H},r)$, which describes the average profile shape, is given by
\begin{equation}
    \xi(R_\mathrm{H},t) = \frac{1}{(2\pi)^3}\int \frac{\mathrm{d}k}{k}\frac{\sin(k r)}{k r} \mathcal{P}_\delta(R_\mathrm{H},k),
    \label{eqn7:profileFunction}
\end{equation}
where equation \eqref{eqn7:powerSpectra} relates the density and curvature power spectra.
We will consider several forms for the curvature perturbation power spectrum, which cover a wide range of models in the literature:
\begin{itemize}

\item the (Dirac-)delta power spectrum
\begin{equation}
    \mathcal{P}_\zeta = \mathcal{A}\delta_\mathrm{D}(k-k_*),
\end{equation}
where $\delta_\mathrm{D}$ is the Dirac-delta function.
It is notable that this doesn't correspond to a realistic or physical power spectrum, it is often used as a reasonable approximation to sufficiently narrow spectra.

\item the narrow-peak power spectrum
\begin{equation}
    \mathcal{P}_\zeta = \mathcal{A}\exp\left( -\frac{(k-k_*)^2}{2\Delta^2} \right),
\end{equation}
which contains a Gaussian-shaped peak. It is also possible to utilise a lognormal peak instead.

\item The power-law power spectrum
\begin{equation}
    \mathcal{P}_\zeta = \mathcal{A}\left(\frac{k}{k_*}\right)^{n_s-1}\Theta_H(k-k_{min})\Theta_H(k_*-k),
\end{equation}
where the Heaviside functions provide both a UV and IR cut-off. We will here consider the scale-invariant power spectrum, $n_s=1$.

\end{itemize}
Whilst these describe very different power spectra, one thing that they all share in common is a cut-off at some scale in both the UV and IR. Without the UV cut-offs, the integral in equation \eqref{eqn7:profileFunction} diverges - and, as we will see later, the variance of the distribution also diverges. The IR cut-off is not so significant, as the $k^4$ term when converting to the density power spectrum, as in equation \eqref{eqn7:powerSpectra}, naturally provides a sharp IR cut-off.

\begin{figure}
\centering
\includegraphics[width=0.95\columnwidth]{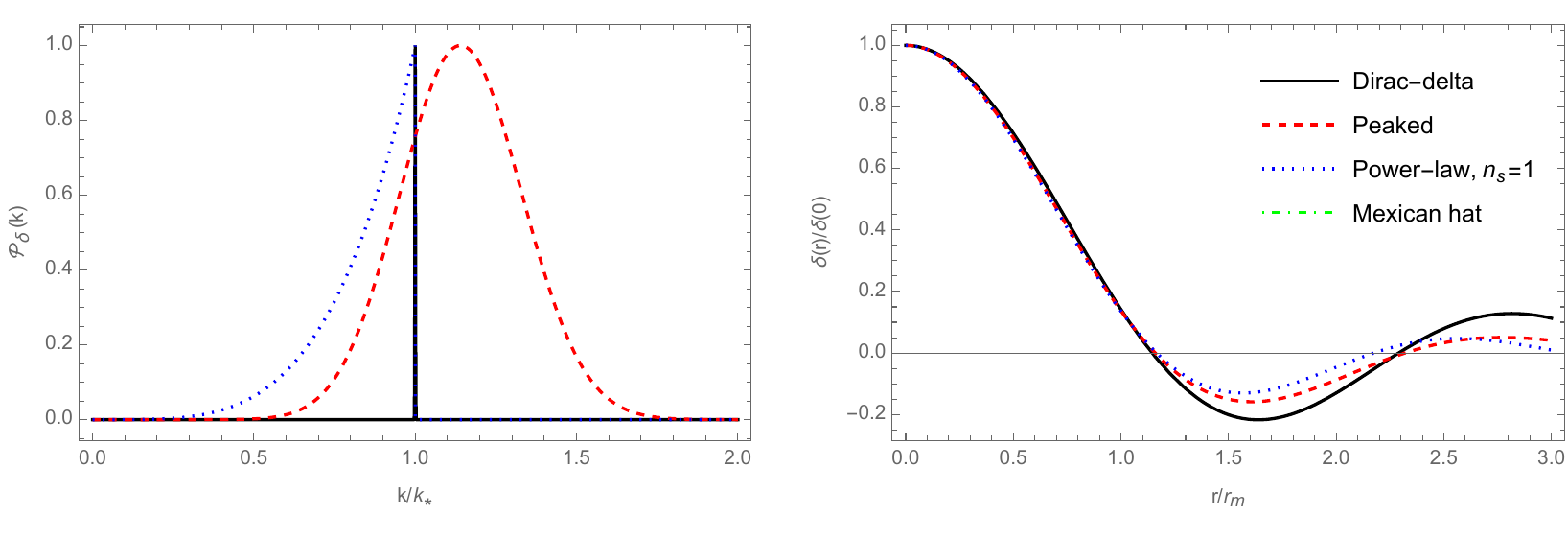}
\caption{\emph{Power spectra and average profile shape.} The left plot shows several different possible choices for the density power spectrum, normalised to have an amplitude of unity at the peak value. The right plot shows that the average profile associated with such power spectrum, with the Mexican hat profile shape also included with the green dotted line.}
\label{fig7:profiles}   
\end{figure}

Figure \ref{fig7:profiles} shows the resulting density power spectra, using equation \eqref{eqn7:powerSpectra}, and average density profiles, calculated using equation \eqref{eqn7:averageProfile}. We can see that, whilst the curvature perturbation power spectra are very different, the density power spectrum are well characterised by a sharp peak. As a result, the average density profile in each case is also very similar (although note that the characteristic scale $r_m$ of the average profiles does vary slightly, and it has been normalised to 1 in the figure), and matches closely to the Mexican-hat profile already considered, equation \eqref{eqn7:mexicanHat}. Because of this, the threshold amplitude for collapse also shows little variation, and we can safely use the threshold for the Mexican-hat profile shape, $C_c\approx 0.5$. There is, however, some dependance on the shape of the power spectrum, which is discussed in reference~\cite{Germani:2018jgr}.

We note that this critical value corresponds to a top-hat smoothing, and, as we will see later, the choice of smoothing function can have a significant impact on the calculated PBH abundance. If the choice is made to use a Gaussian smoothing function for the calculation, then the critical value for the Gaussian-smoothed compaction should be used: $C_{G,c} \approx 0.18$. When properly accounted for, this introduces an uncertainty in the amplitude of the power spectrum required to form a given abundance of PBHs of order $10\%$~\cite{Young:2019osy}.

\section{Methods for the calculation}
\label{sec7:methods}

We will first discuss some generalities with the calculation, before discussing the simplest method, a \emph{Press-Schechter}-type approach, and finally the theory of peaks and its extensions. Throughout this chapter, we will assume a Gaussian distribution for the curvature perturbation $\zeta$. At first-order, the density $\delta$, compaction $C$ and curvature perturbation $\zeta$ are linearly related, implying both $\delta$ and $C$ also follow a Gaussian distribution. Since we will use $C$ for the calculation, we will consider the probability density function (PDF) of $C$ given by,
\begin{equation}\label{eqn7:P(C)}
    P(C) = \frac{1}{\sqrt{2\pi \sigma_0^2}}\exp\left( -\frac{C^2}{2\sigma_0^2} \right),
\end{equation}
where $\sigma_0^2$ is its variance. The variance can be calculated by taking the zeroth-order (n=0) moment of the power spectrum,
\begin{multline}
    \sigma_n^2(R_s,R_\mathrm{H}) = \int\limits_0^\infty \frac{\mathrm{d}k}{k}k^{2n}\mathcal{P}_C(R_s,R_\mathrm{H},k)\\
    = \int\limits_0^\infty \frac{\mathrm{d}k}{k}k^{2n}\frac{4(1+\omega)^2}{(5+3\omega)^2} \left(k R_s \right)^4\mathcal{P}_\zeta(k)\tilde{W}^2(R_s,k)T^2(R_\mathrm{H},k),
    \label{eqn7:momentsP}
\end{multline}
where $R_S$ is the smoothing scale, and $R_\mathrm{H}$ is the Hubble horizon. For the \emph{Press-Schechter}-type approach, we will only need $\sigma_0^2$, whilst higher order terms will be needed for the theory of peaks. 

Several things are note-worthy at this point. The first is that, whilst PBH formation is an extremely non-linear process, $T(R_\mathrm{H},k)$ is the linear transfer function. It is, therefore, preferable to complete calculations in the super-horizon limit $R_s\ll R_\mathrm{H}$, such that $\delta\ll 1$, and $T(R_\mathrm{H},k)\approx1$. The second is that the integrand contains the factor $k^4$, which can cause the integral to diverge if there is no UV cut-off sharper than $k^{-4}$. This is the case for a Gaussian smoothing, equation \eqref{eqn7:fourierGaussian}, which has an exponential cut-off, but not necessarily for the top-hat smoothing, equation \eqref{eqn7:fourierTH}, which (though highly oscillatory) drops as $k^{-4}$. Therefore, when the top-hat function is used, the calculation is often performed at horizon crossing, $R_S=R_\mathrm{H}$ on a somewhat \emph{ad hoc} basis, and the effect of the transfer function then guarantees that the integral converges. For further reading on this topic, it is discussed in more detail in~\cite{Young:2019osy}, and the effect of non-linearities at horizon crossing has been studied in~\cite{DeLuca:2023tun}.

The initial abundance of PBHs is typically stated in terms of $\beta$, the energy fraction of the universe collapsing into black holes at a given time $t$,
\begin{equation}
    \beta(t) = \left. \Omega_\mathrm{PBH}(t) \right|_\mathrm{formation} = \left. \frac{\rho_\mathrm{PBH}}{\rho_r} \right|_\mathrm{formation}, 
\end{equation}
where $\rho_\mathrm{PBH}$ is the mean energy density of PBHs, $\rho_r$ is the mean radiation density, and we are assuming complete radiation domination. It will be helpful for consider $\beta$ as the fraction of the universe collapsing to form PBHs at a time $t$ when the Hubble horizon is $R_s$ and the horizon mass is $M_\mathrm{H}$, where these 3 parameters can be used interchangeably. Note that the smoothing scale $R_s$ is used here, rather than the (actual) Hubble horizon $R_\mathrm{H}$ at the time of calculation. This is because the calculation typically works by comparing the \emph{initial} amplitude of perturbations to the critical amplitude, long before such perturbations enter the horizon. The scale of the perturbations forming PBHs, corresponding to the smoothing scale, is therefore much larger than the Hubble scale at this time. However, as discussed in the previous paragraph, it is often necessary to set $R_s=R_\mathrm{H}$.

Throughout this chapter, we will assume that PBHs form spontaneously upon horizon re-entry. Accounting for the fact that PBHs take some time to form after horizon re-entry would mean a larger value for $\beta$ (because $\rho_r$ drops faster than $\rho_\mathrm{PBH}$ as the universe expands), which would then be cancelled out by the relative changes in the PBH and radiation densities as the universe expands when the final PBH abundance is calculated at a later time. For simplicity, this is therefore neglected here, but may be important if one is considering the formation of very light PBHs which evaporate extremely rapidly.

\subsection{The \emph{Press-Schechter}-type approach}
\label{sec7:PS}

The idea behind the \emph{Press-Schechter}-type approach is that we can calculate $\beta$, by calculating the fraction of the volume of the universe where the compaction is above the threshold value. This gives us the collapse constraint
\begin{equation}
    n(\bar{\nu}) = \delta_\mathrm{D}\left( \nu - \bar{\nu} \right)\Theta_H(\nu-\nu_c),
\end{equation}
where $\delta_\mathrm{D}$ is the Dirac-delta function, $\nu_c = C_c/\sigma_0$, and 
\begin{equation}
    \nu(\mathbf{x}) = \frac{C(\mathbf{x})}{\sigma_0},
    \label{eqn7:nu}
\end{equation}
has unit variance. This returns the number of regions (either 0 or 1) where $\nu$ takes a value between $\bar{\nu}$ and $\bar{\nu}+\mathrm{d}\nu$, above the critical value at point $\mathbf{x}$. To calculate $\beta$, we will want the expectation value of $n$:
\begin{equation}
    \mathcal{N}(\bar{\nu}) = \langle n \rangle = \int\limits_{-\infty}^\infty\mathrm{d}\nu n(\bar{\nu}) P(\nu) = 
    \begin{cases}
        P(\bar{\nu}) \quad \quad \bar{\nu}\geq \nu_c \\
        0 \quad \quad \quad   \bar{\nu}< \nu_c \\
    \end{cases}
    .
\end{equation}
We can then calculate $\beta$ as
\begin{equation}
    \beta(t) = 2 \int\limits_{\nu_c}^{\infty}\mathrm{d}\nu\frac{M_\mathrm{PBH}}{M_\mathrm{H}}\frac{1}{\sqrt{2\pi}}\exp \left(-\frac{\nu^2}{2} \right),
    \label{eqn7:betaPS}
\end{equation}
where the fraction $M_\mathrm{PBH}/M_\mathrm{H}$ gives the fraction of each Hubble patch which collapses to form a PBH. The time dependance is somewhat hidden here, 

The simplest scenario is to assume that PBHs form with a fixed fraction of the horizon mass, $M_\mathrm{PBH}/M_\mathrm{H}=\alpha$, whilst simulations of PBH formation show that the PBH mass depends on both the scale and amplitude of the perturbation from which it formed, which will be discussed in section \ref{sec7:criticalScaling}. The factor 2 arises as a ``fudge-factor'' from requiring that, for $\nu_c\ll 1$, all of the energy universe should collapse to form PBHs, $\beta=1$. Without the addition of the factor of 2, this integral would give $\beta=0.5$ (for $\alpha=1$).

For $M_\mathrm{PBH}/M_\mathrm{H}=\alpha$, the expression for $\beta$ is simply the complementary error function $\mathrm{erfc}$,
\begin{equation}
    \beta(t) = \alpha \mathrm{erfc}\left( \frac{\nu_c}{\sqrt{2}} \right).
\end{equation}
Using the large $x$ limit of $\mathrm{erfc}(x)$ gives,
\begin{equation}
    \beta(t) \approx \frac{\alpha}{\sqrt{2\pi}\nu_c}\exp\left(-\frac{\nu_c^2}{2} \right).
    \label{eqn7:betaApproximation}
\end{equation}
Whilst not very accurate, this analytic expression does give an order-of-magnitude estimate for the amplitude of the power spectrum required to form a given abundance of PBHs, and illustrate the exponential dependence of $\beta$ on $\sigma_0$ (and thus, on the power spectrum). Because PBHs form from rare perturbations in the extreme tail of the distribution, $\nu_c\gg 1$, small changes in the power spectrum result in large changes to $\beta$.

\subsubsection{The critical scaling relationship}
\label{sec7:criticalScaling}

When a perturbation collapses to form a PBH, the final mass depends on how much of the surrounding matter has been pulled in. For this to happen, gravitational attraction towards the centre needs to overcome pressure forces and the initial Hubble expansion of the surrounding material. The larger the density at the centre, the greater the force of gravity, and the more material is pulled into the final PBH. The result is that PBH mass depends on the amplitude of the perturbation from which it formed, as well as the scale. Simulations of PBH formation reveal that the PBH follows a power law known as the critical scaling relation~\cite{Musco:2008hv,Musco:2012au,Young:2019yug},
\begin{equation}
    M_\mathrm{PBH} = \mathcal{K}M_\mathrm{H}\left( C-C_c \right)^\gamma,
    \label{eqn7:criticalScaling}
\end{equation}
where the exact values depend on the specific profile considered, but we will take $\mathcal{K}=4$, and $\gamma=0.36$ as representative values during radiation domination. The horizon mass when the comoving Hubble horizon is $R_H$, assuming radiation domination until the time of matter-radiation equality, is given by
\begin{equation}
    M_\mathrm{H}(R_\mathrm{H}) = M_\mathrm{eq}\left( \frac{g_*}{g_\mathrm{\star,eq}} \right)^{-1/6}\left( \frac{R_\mathrm{H}}{R_\mathrm{eq}} \right)^2 M_\mathrm{eq} \approx 0.5 \left( \frac{g_*}{10.75} \right)^{-1/6}\left( \frac{R_\mathrm{H}}{10^{-6} \mathrm{Mpc}} \right)^2M_\odot,
    \label{eqn7:horizonMass}
\end{equation}
where $g_*$ is the number of relativistic degrees of freedom (although the dependance on this is extremely weak and is typically neglected), the subscript $\mathrm{eq}$ denotes the value at matter-radiation equality, and $M_\odot$ is a solar mass. For the parameter choices in~\cite{Nakama:2016gzw}, $M_\mathrm{eq} = 2.8\times 10^{17}M_\odot$. 
Here, we have given the horizon mass $M_\mathrm{H}$ as a function of the Hubble horizon $R_H$ at a given time, but for our purposes it will be more convenient to consider what the horizon mass will be at the time when the smoothing scale enters the horizon, which is achieved simply by replacing $R_\mathrm{H}$ with $R_s$. Therefore, the expression we will use for the horizon mass is therefore
\begin{equation}
    M_\mathrm{H}(M_\mathrm{H}) \approx 0.5 \left( \frac{g_*}{10.75} \right)^{-1/6}\left( \frac{R_s}{10^{-6} \mathrm{Mpc}} \right)^2M_\odot.
    \label{eqn7:smoothingMass}
\end{equation}

Utilising equation \eqref{eqn7:criticalScaling}, the expression for the initial PBH abundance becomes
\begin{equation}
    \beta(t) = 2 \int\limits_{\nu_c}^{\infty}\mathrm{d}\nu \mathcal{K}\left(\nu\sigma_0-\nu_c\sigma_0 \right)^\gamma \frac{1}{\sqrt{2\pi}}\exp \left(-\frac{\nu^2}{2} \right),
    \label{eqn7:betaPSscaling}
\end{equation}
where the integral is now performed numerically. 

Whilst not needed at this point, it will be helpful later to express $\beta$ as an integral over the PBH mass. Equations \eqref{eqn7:criticalScaling} and \eqref{eqn7:nu} can be used to express $\nu$ as a function of the PBH mass
\begin{equation}
    \nu(M_\mathrm{PBH}) = \frac{C(M_\mathrm{PBH})}{\sigma_0} = \frac{1}{\sigma_0}\left(\frac{M_\mathrm{PBH}}{\mathcal{K}M_\mathrm{H}}\right)^{1/\gamma}+\nu_c,
    \label{eqn7:nu(M)}
\end{equation}
and differentiated to give
\begin{equation}
    \frac{\mathrm{d}\nu}{\mathrm{d}M_\mathrm{PBH}} = \frac{1}{\gamma \mathcal{K}M_\mathrm{H}\sigma_0} \left(\frac{M_\mathrm{PBH}}{\mathcal{K}M_\mathrm{H}}\right)^{\frac{1}{\gamma}-1}.
\end{equation}
Performing a change of variable, we can then express $\beta$ as
\begin{equation}
    \beta(t) = 2\int\limits_0^\infty \mathrm{d}M_\mathrm{PBH}\left[ \frac{1}{\gamma \mathcal{K}M_\mathrm{H}\sigma_0} \left(\frac{M_\mathrm{PBH}}{\mathcal{K}M_\mathrm{H}}\right)^{\frac{1}{\gamma}-1} \right] \frac{M_\mathrm{PBH}}{M_\mathrm{H}}\frac{1}{\sqrt{2\pi}}\exp \left(-\frac{\nu^2(M_\mathrm{PBH})}{2} \right).
    \label{eqn7:betaMpbh}
\end{equation}

\begin{figure}
\centering
\includegraphics[width=0.7\columnwidth]{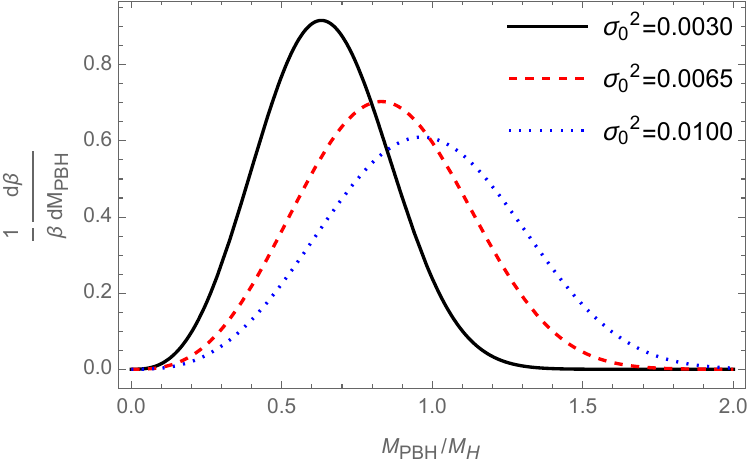}
\caption{\emph{Critical scaling and extended mass functions.} The figure shows the masses of PBHs forming at a specific time (or, equivalently, at a specific scale), normalised by the total abundance. The critical scaling relationship between the PBH mass and the amplitude of perturbations means that PBHs form at a range of masses, even when they only form at one specific time. The larger the amplitude of perturbations, the larger the mass of the PBHs which form.}
\label{fig7:extendedMass}   
\end{figure}

Including the effect of the critical scaling relationship means that, even when only considering perturbations of a single scale (i.e. PBHs forming at a single time), PBHs form with a range of masses. It also means that, in addition to affecting the total abundance of PBHs which form, the amplitude of the power spectrum also affects the masses at which they form. Figure \ref{fig7:extendedMass} shows the abundance of PBH forming at different scales, normalised by their overall abundance, with values for the variance $\sigma_0^2=\{0.003,0.0065,0.01$\}. These were chosen to correspond approximately with typical values for the abundance, and give $\beta\approx \{4\times 10^{-20},4\times 10^{-10},5\times 10^{-7}\}$. For larger $\sigma_0^2$, the distribution peaks at larger masses - although still very similar to the horizon mass. NB. This plot is independent of the shape of the power spectrum. In the \emph{Press-Schechter} formalism, the power spectrum only enters into the calculation via the $\sigma_0^2$ term - which we have here simply chosen values for.

\subsubsection{The total primordial black hole abundance}
\label{sec7:totalAbundance}

We have so far considered the abundance of PBHs at a single time of formation, $\beta$. In order to calculate the total abundance of PBHs at a later time (e.g. today), we need to sum the amount of PBHs forming at all times, and account for the evolution of their density. Assuming that the universe behaves purely as radiation domination until the time of matter-radiation equality, then we can express the abundance of PBHs which formed when the horizon mass was $M_\mathrm{H}$ as a function of $\beta$,
\begin{equation}
    \beta_\mathrm{eq}(M_\mathrm{H}) = \frac{a_\mathrm{eq}}{a(M_\mathrm{H})}\beta(M_\mathrm{H}) = \left(\frac{M_\mathrm{eq}}{M_\mathrm{H}}\right)^{1/2}\beta(M_\mathrm{H}).
\end{equation}
The total PBH abundance at matter-radiation equality is therefore found by integrating over all times (scales) at which PBHs form,
\begin{equation}
    \Omega_\mathrm{PBH} = \int\limits_{M_\mathrm{min}}^{M_\mathrm{max}} \frac{\mathrm{d}M_\mathrm{H}}{M_\mathrm{H}}\left(\frac{M_\mathrm{eq}}{M_\mathrm{H}}\right)^{1/2}\beta(M_\mathrm{H}).
    \label{eqn7:omegaPBH}
\end{equation}
The integral should technically be performed over all scales, but in practice it is only necessary between the minimum and maximum scales being considered. Typically, this range is set by the power spectrum under consideration, where PBH formation occurs over a limited range of scales near the peak of the power spectrum (or, between UV and IR cut-offs). 

The total PBH abundance is often also expressed in terms of the fraction of (cold) dark matter composed of PBHs
\begin{equation}
    f_\mathrm{PBH} = \frac{\Omega_\mathrm{PBH}}{\Omega_\mathrm{CDM}},
    \label{eqn7:fPBH}
\end{equation}
where $\Omega_\mathrm{CDM}$ is the dark matter density parameter. This is also useful because it is a time-independent quantity (after PBHs have stopped forming).

\subsubsection{The mass function}

In addition to the total abundance, a key potential observable of PBHs is the mass function $f_\mathrm{PBH}$, which describes how many PBHs there are at different masses. There are several different ways in which this can be parameterised, but we will use the definition,
\begin{equation}
    \psi\left( M_\mathrm{PBH} \right) = \frac{1}{\Omega_\mathrm{CDM}}\frac{\mathrm{d}\Omega_\mathrm{PBH}}{\mathrm{d} (\ln M_\mathrm{PBH})} = M_\mathrm{PBH} \frac{\mathrm{d}f_\mathrm{PBH}}{\mathrm{d} M_\mathrm{PBH}},
\end{equation}
which satisfies the normalisation condition,
\begin{equation}
    \int\mathrm{d}(\ln M_\mathrm{PBH})\psi(M_\mathrm{PBH}) = f_\mathrm{PBH}.
\end{equation}

Using equations \eqref{eqn7:betaMpbh}, \eqref{eqn7:omegaPBH} and \eqref{eqn7:fPBH} gives us the final expression for the mass function
\begin{multline}
    \psi(M_\mathrm{PBH}) = \frac{1}{\Omega_\mathrm{CDM}}\int\limits_{M_\mathrm{min}}^{M_\mathrm{max}} \frac{\mathrm{d}M_\mathrm{H}}{M_\mathrm{H}}\left(\frac{M_\mathrm{eq}}{M_\mathrm{H}}\right)^{1/2} \frac{2}{\gamma \mathcal{K}M_\mathrm{H}\sigma_0} \left(\frac{M_\mathrm{PBH}}{\mathcal{K}M_\mathrm{H}}\right)^{\frac{1}{\gamma}-1}  \\
    \times \frac{M_\mathrm{PBH}}{M_\mathrm{H}}\frac{1}{\sqrt{2\pi}}\exp \left(-\frac{\nu^2(M_\mathrm{PBH})}{2} \right),
    \label{eqn7:psi}
\end{multline}
where the expression for $\nu(M_\mathrm{PBH})$ is given in equation \eqref{eqn7:nu(M)}. Recall that the variance $\sigma_0^2$ is a function of the smoothing scale, which gives the integrand a dependence on the horizon mass $M_\mathrm{H}$.

\subsection{The theory of peaks}
\label{sec7:BBKS}

The theory of peaks, developed by Bardeen, Bond, Kaiser, and Szalay (BBKS) \cite{Bardeen:1985tr}, extends upon the \emph{Press-Schechter} formalism by stating that compact objects (PBHs in this case) will form at peaks in the density above some threshold value. Our collapse constraint becomes
\begin{equation}
    n(\mathbf{x}) = \delta_D^{(3)}(\mathbf{\eta})\Theta_H(\lambda_3) \delta_\mathrm{D}^{(3)}\left( \nu - \bar{\nu} \right)\Theta_H(\nu-\nu_c),
\end{equation}
where the index $(3)$ indicates a 3-dimensional function, $\sigma_n$ are moments of the power spectrum given by equation \eqref{eqn7:momentsP}, $\mathbf{\eta}$ is given by
\begin{equation}
    \mathbf{\eta}(\mathbf{x}) =\frac{1}{\sigma_1}\mathbf{\nabla}C(\mathbf{x}) 
\end{equation}
$\zeta_{ij}$ is given by
\begin{equation}
    \zeta_{ij}(\mathbf{x})=\frac{1}{\sigma_2}\frac{\partial^2C(\mathbf(x))}{\partial x_i \partial x_j},
\end{equation}
and finally, $\lambda_3$ is the smallest eigenvalue of $-\zeta_{ij}$. The collapse constraint $n(\mathbf{x})$ returns the number of peaks (either 0 or 1) of height between $\bar{\nu}$ and $\bar{\nu}+\mathrm{d}\nu$, above the critical value $\nu_c$ in the infinitesimal volume $\mathrm{d}^3x$, divided by $\mathrm{d}^3x$. This is analogous to the collapse constraint used in the \emph{Press-Schechter}-type approach, where the additional $\delta_D^{(3)}(\mathbf{\eta})$ term constrains the spatial derivative of $C$ to be zero, whilst the $\Theta_H(\lambda_3)$ term specifies peaks rather than troughs. 

The collapse constraint depends on the spatial coordinate $\mathbf{x}$ only through a set of continuous Gaussian fields evaluated at the same position, which we will denote $\mathbf{y}$,
\begin{equation}
    \mathbf{y} = \{ \nu, \eta_i, \zeta_{ij} \}.
\end{equation}
In order to calculate the abundance of PBHs, we need to calculate the average value for $n$, calculated by integrating over the PDF $P(\mathbf{y})$
\begin{equation}
    \mathcal{N} = \int \mathrm{d}\mathbf{y}\frac{\sigma_2^3}{\sigma_1^3}\left| \mathrm{det}\zeta_{ij} \right|n P(\mathbf{y}).
\end{equation}
The term $\frac{\sigma_2^3}{\sigma_1^3}\left| \mathrm{det}\zeta_{ij} \right|$ is included to account for the transformation from the Gaussian variables of the constraint to the physical coordinates $\mathbf{x}$.
Whilst not straightforward to calculate, the integral gives the result
\begin{multline}
    \mathcal{N}(\nu)=\frac{\sigma_2^3}{3^{3/2}(2\pi)^2\sigma_1^3}\exp\left(-\frac{\nu^2}{2}\right)\Theta_H(\nu-\nu_c)\\
    \times\int\limits_0^\infty\mathrm{d}u f(u) \frac{1}{\sqrt{2\pi}(1-\gamma_{02}^2)}\exp\left( -\frac{(u-\gamma_{02}\nu)^2}{2(1-\gamma_{02}^2)}  \right),
    \label{eqn7:fullpeaks}
\end{multline}
where $\gamma_{02}=\sigma_1^2/\sigma_0\sigma_2$ is the correlation function, and the function $f(u)$ is given by
\begin{multline}
    f(u) = \sqrt{\frac{2}{5\pi}}\left[\left( \frac{u^2}{2} - \frac{8}{5} \right)\exp\left( -\frac{5u}{2} \right) + \left( \frac{31 u^2}{4} + \frac{8}{5} \right)\exp\left( -\frac{5u}{8} \right)  \right] \\
    + \frac{1}{2}\left( u^3-3u \right)\left[ \mathrm{erf}\left( \sqrt{\frac{5}{2}}u \right)+\mathrm{erf}\left( \sqrt{\frac{5}{2}}\frac{u}{2} \right) \right],
    \label{eqn7:f}
\end{multline}
where $\mathrm{erf}(x)$ is the error function
\begin{equation}
    \mathrm{erf}(x) = \frac{2}{\pi}\int\limits_0^x\mathrm{d}t\exp(-t^2).
\end{equation}
For the full derivation, the interested reader is directed towards the original BBKS paper~\cite{Bardeen:1985tr}, as well as section II and III of~\cite{Lazeyras:2015giz}.

This expression is somewhat complicated, but fortunately can be greatly simplified when one considers that PBHs form from high peaks, and one can then make the assumption $v \gg 1$, which is known as the high-peak limit. The integrand appearing in equation \eqref{eqn7:fullpeaks} is the function $f(u)$ multiplied by a Gaussian PDF with mean $\gamma_{02}\nu$ and variance $(1-\gamma_{02}^2)$, where we generically have $\gamma_{02}\nu \gg (1-\gamma_{02}^2)$. 
This means that the PDF can be well approximated by a Dirac-delta function, $\delta_D(u-\gamma_{02}\nu)$. Under this approximation, the integral is then equal to $f(\gamma_{02}\nu)$, with $\gamma_{02}\nu\gg 1$. The exponential terms in equation \eqref{eqn7:f} are then very small, whilst the $\mathrm{erf}$ are close to unity. We then have
\begin{equation}
    f(u) \approx u^3-3u \approx u^3.
\end{equation}
Inserting this into equation \eqref{eqn7:fullpeaks} gives the us the well-known (and much simpler) number density of peaks in the high-peak limit:
\begin{equation}
    \mathcal{N}_{hi-pk}(\nu) = \frac{1}{3^{3/2}(2\pi)^2}\frac{\sigma_1^3}{\sigma_0^3}\exp\left( -\frac{\nu^2}{2} \right).
\end{equation}
From this expression, we can easily that the factor $\sigma_1^3/\sigma_0^3$ gives the expression units of $k^3 \propto 1/\mathrm{volume}$, since $\mathcal{N}$ is a number density (rather than simply a number, as in section \ref{sec7:PS}).

We can now proceed with the calculation of the PBH abundance, which proceeds in much the same way as before, except that, since $\mathcal{N}$ now gives a number density, we now need to multiply by a volume to find the fraction of the universe collapsing into PBHs at a given time (or equivelently, scale or horizon mass):
\begin{equation}
    \beta(R_s) = V(R_s) \int\limits_{\nu_c}^\infty \mathrm{d}\nu \frac{M_\mathrm{PBH}}{M_\mathrm{H}}\frac{1}{3^{3/2}(2\pi)^2}\frac{\sigma_1^3}{\sigma_0^3}\exp\left( -\frac{\nu^2}{2} \right),
\end{equation}
where, again, the critical scaling relationship, equation \eqref{eqn7:criticalScaling}, can be used to relate the PBH mass to the amplitude of the perturbation which formed it. The volume term $V(R_s)$ is the volume of the window function being used,
\begin{equation}
    V(R_s)=
    \begin{cases}
        (2\pi)^{3/2}R_s^3 \quad \quad  \mathrm{Gaussian ~smoothing}, \\
        \frac{4\pi}{3}R_s^3 \quad \quad \quad \quad \mathrm{Top-hat~ smoothing}.  \\
    \end{cases}
\end{equation}
The total abundance and mass function can then be calculated using the same method as in section \ref{sec7:PS}. 

It is noteworthy that the dominant term in the expression for $\beta$ remains the unchanged exponential term. However, using the theory of peaks, rather than \emph{Press-Shechter} generally predicts a higher abundance of PBHs, due to the additional $v^3$ term appearing in the expression. However, when one calculates the amplitude of the power spectrum required to produce a specified PBH abundance, the difference is typically only of order $10\%$, although significant changes can be seen in the mass function~\cite{Gow:2020bzo,Young:2014ana}.

\subsection{Extensions to the theory of peaks}
\label{sec7:YM}

When studying the large-scale structure of the universe, a similar approach is taken to that considered here. Initial fluctuations in the density of the universe grow over time and provide the seeds for the formation of compact objects (such as galaxies and galaxy clusters, in this case), and one can calculate the later abundance of e.g. galaxies by considering the initial distribution of perturbations and the amplitude required for collapse. A common problem encountered in this case is the cloud-in-cloud phenomenon, which arises when a small-scale compact object is calculated to form inside a large-scale compact object, resulting in double counting. A common method used to solve this problem for large-scale structure is the excursion set approach, which works by not adding the stipulation that haloes are only considered to collapse to form a compact object if there are no haloes on larger scales which will form compact objects at that scale. 

The excursion set has been considered in the case of PBH formation, and it is found that the cloud-in-cloud phenomenon, whereby a smaller mass PBH would be absorbed by bigger mass PBHs forming later, is basically absent for a Gaussian distribution~\cite{MoradinezhadDizgah:2019wjf}. This is due to the rarity of PBHs at formation, meaning that only a very small fraction of small mass PBHs form at locations where larger mass PBHs later form. \footnote{However, this may be different for non-Gaussian distributions, which could mean that PBHs form at small masses at a location where there is a large-scale perturbation which will collapse to a large mass PBH (although this does not occur for non-Gaussianities arising as a result of non-linearities, which are considered in section \ref{sec7:non-linearities}).} 

However, a related problem to the cloud-in-cloud problem can be relevant for PBHs, which is most easily seen if we consider a single perturbation, such as that shown in figure \ref{fig7:perturbation}. For a smoothing scale $R_s=r=0$, the compaction $C$ is also zero, and as the smoothing scale increases, $C$ peaks at the characteristic scale $r_m$ before dropping back to zero. If the peak value of $C$ is above the threshold for collapse, then the perturbation will collapse to form a PBH, with a mass determined by the amplitude of $C$ at the peak and the scale $r_m$ (as in equation \eqref{eqn7:criticalScaling}). However, this does mean that the compaction is above the threshold for a range of smoothing scales (unless the peak occurs at exactly $C=C_c$, in which case a vanishingly small PBH forms). The calculations that we have considered thus far would, therefore, consider that PBHs form at each of these scales, and sum over the relevant range of scales. However, in reality, only a single PBH would form, corresponding to the scale at which $C$ peaks.

This can be accounted for by extending the theory of peaks to stipulate that PBHs form not only at locations where the compaction peaks, but also at the (smoothing) scale at which the compaction peaks:
\begin{equation}
    n(\mathbf{x}) = \delta_D\left(\frac{\partial C}{\partial R}\right)\Theta_H\left(\frac{\partial ^2C}{\partial C^2} \right) 
    \delta_D^{(3)}(\mathbf{\eta})\Theta_H^{(3)}(\lambda_3) 
    \delta_\mathrm{D}^{(3)}\left( \nu - \bar{\nu} \right)\Theta_H(\nu-\nu_c).
\end{equation}
The calculations necessary were developed simultaneously by Germani \& Sheth~\cite{Germani:2019zez} and Young \& Musso~\cite{Young:2020xmk}. The effect of accounting for the improved the theory of peaks calculated was considered by Gow et al~\cite{Gow:2020bzo}, and it was found that the change to power spectrum amplitude to produce a given PBH abundance only varies by a few percent, and that the calculated mass function is also largely unchanged. For this reason, the extensions to the theory of peaks are typically not included in calculations.

\section{The effect of non-linearities}
\label{sec7:non-linearities}

Until this point, we have been considering a linear relation between the curvature perturbation $\zeta$ and the density contrast $\delta$, resulting also in a linear relation to the compaction $C$. However, for PBHs to form, large amplitude perturbations are needed, and it is therefore necessary to go beyond linear order~\cite{Young:2019yug}. 

The first effect we need to account for is a coordinate transform between the comoving uniform density gauge in which we have defined $\zeta$ and the comoving synchronous gauge in which we calculate $C$ and determine PBH formation. We will still consider that PBHs form at high peaks which are spherically symmetric, and therefore only consider the radial coordinate. The radial coordinate $\hat r$ in the comoving synchronous gauge is then related to the radial coordinate $r$ appearing in the comoving uniform density gauge by
\begin{equation}
    \hat r = a(t)\exp\left( \zeta( r) \right)r.
\end{equation}
For small amplitude perturbations, $\hat r$ increases monotonically with $r$, and these are known as type I perturbations. However, as larger $\zeta$ is considered, there comes a point where this is longer true, meaning that increasing coordinate radius $r$ does not coincide with increasing physical distance. Perturbations where this occurs are known as type II perturbations, and their evolution (and their subsequent collapse to PBHs) is not currently well understood. We can find the point at which the switch between type I and II perturbations occurs by searching for a stationary point $\mathrm{d}\hat r/\mathrm{d} r =0$. Stationary points occur when we have
\begin{equation}
    -r\zeta'(r) = 1.
\end{equation}
For any given perturbation, this function peaks at the scale $r_m$ (see the discussion in section \ref{sec7:profiles}), and we therefore have the condition for type II perturbations,
\begin{equation}
    -r_m\zeta'(r_m) > 1.
    \label{eqn7:type2}
\end{equation}

The second effect we need to consider is the non-linear relationship between $\zeta$ and $\delta$, and we will again assume that $\zeta$ follows a Gaussian distribution. In the super-horizon regime, $\delta$ can be expressed as a function of $\zeta$ as
\begin{equation}
    \delta(\hat r) = -\frac{4(1+\omega)}{5+3\omega}\left(\frac{1}{aH}\right)^2\exp\left( -\frac{5\zeta(r)}{2} \right)\nabla^2\exp\left( \frac{\zeta(r)}{2} \right),
    \label{eqn7:deltaNL}
\end{equation}
where $\hat r$ is the radial coordinate in the comoving synchronous gauge, and in spherical symmetry
\begin{equation}
    \nabla^2 = \frac{\partial^2}{\partial r^2}+\frac{2}{r}\frac{\partial}{\partial r}.
\end{equation}
We note that equation \eqref{eqn7:deltaNL} is exact to all orders in $\zeta$, but is correct to second order in $\epsilon=1/HL$, where $L$ is the length scale characterizing the perturbation, and this expression is therefore valid in the super-horizon regime.

Combining these equations and inserting them into the equation for the compaction with a top-hat window function, equation \eqref{eqn7:Cth} gives
\begin{multline}
    C(r) = -\frac{2}{a e^{\zeta(r)}r}\frac{3H^2}{8\pi}\int\limits_0^r \mathrm{d}\left( a e^{\zeta(r)}r \right) \left[ 4\pi \left(a e^{\zeta(r)}r\right)^2 \right]\\
    \times \frac{2(1+\omega)}{5+3\omega}\left(\frac{1}{aH}\right)^2 e^{-2\zeta(r)}\left[ \zeta''(r)+\frac{2}{r}\zeta'(r)+\frac{1}{2}\left( \zeta'(r) \right)^2 \right],
\end{multline}
where the prime denotes a derivative with respect to $r$.
Fortunately, this complicated looking integral has a simple analytic solution:
\begin{equation}\label{eqn7:C-zeta'}
    C(r) = -\frac{3(1+\omega)}{5+3\omega}\left[2r\zeta'(r) -\left(r\zeta'(r)\right)^2\right].
\end{equation}
The linear component of this expression,
\begin{equation}
    C_1(r) = \frac{6(1+\omega)}{5+3\omega}r\zeta'(r),
\end{equation}
is exactly the same as the expression for the compaction seen in equation \eqref{eqn7:C1zeta}, and the variance $\langle C_1^2\rangle = \sigma_0^2$ is calculated exactly as before, as in equation \eqref{eqn7:momentsP}. Setting $\omega=1/3$ during radiation domination, we have a simple quadratic relation between the compaction $C$ and its linear component $C_1$:
\begin{equation}
    C = C_1 -\frac{3}{8}C_1^2,
    \label{eqn7:quadraticCompaction}
\end{equation}
which takes a maximum value $C=2/3$ when $C_1=(4/3)r\zeta'(r)=4/3$. If we consider the characteristic scale $r_m$, where the compaction peaks, this corresponds to the switch between type I and II perturbations, seen in equation \eqref{eqn7:type2}. 

For the large, rare perturbations which form PBHs, peaks in the non-linear compaction field will correspond to peaks in the linear compaction field~\cite{Young:2019yug}, and we can therefore simply apply the same calculation methods we have used before, with some small modifications. We will consider that PBHs form for type I perturbations with $C>C_c$. Solving equation \eqref{eqn7:quadraticCompaction} for $C_1$ gives
\begin{equation}
    C_{1\pm} = \frac{4}{3}\left(1 \pm \sqrt{\frac{2-3C}{2}}\right),
\end{equation}
which gives two solutions for any value $C<2/3$, where $C_{1-}$ and $C_{1+}$ correspond to type I  and II perturbations respectively. Going forwards, we will consider only type I solutions, and will drop the minus sign in the subscript. Taking $C_c\approx 0.5$ (for Mexican hat-type perturbations) means that we will consider PBHs to form in the range,
\begin{equation}
    C_{1,c} \approx 0.67 < C < \frac{4}{3}.
\end{equation}
Due to the exponential suppression in the abundance of large peaks, the upper limit will have hardly any effect on the total abundance of PBHs (although we will need to account for it in the calculation nonetheless, to avoid the appearance of imaginary solutions).

Neglecting the upper bound for the moment, therefore, we can obtain a simple estimate for the effect of non-linearities using equation \eqref{eqn7:betaApproximation}, using $C_{1,c}\approx0.67$ instead of $C_c=0.5$. The result is a severe reduction in the number of PBHs forming, typically by numerous orders of magnitude. In order to achieve the same PBH abundance as before, then $\sigma_0^2$ and the amplitude of the power spectrum needs to be larger by a factor,
\begin{equation}
    \left(\frac{C_{1,c}}{C_c}\right)^2 = \left( \frac{4}{3C_c}\left(1 \pm \sqrt{\frac{2-3C_c}{2}}\right) \right)^2 \approx 1.78,
\end{equation}
where we have used $C_c=0.5$ in the final approximation on the right-hand side. This means that, when constraints on the PBH abundance are translated into constraints on the primordial power spectrum, they are almost a factor of 2 weaker than if non-linearities were ignored.

The main differences we will need to make in the calculation of the PBH abundance concerns the critical scaling relationship, and the limits of several integrals. Inserting equation \eqref{eqn7:quadraticCompaction} into the equation for the critical scaling relation, equation \eqref{eqn7:criticalScaling}, gives the PBH mass as a function of $C_1$
\begin{equation}
    M_\mathrm{PBH} = \mathcal{K}M_\mathrm{H}\left(C_1-\frac{3}{8}C_1^2-C_c \right)^\gamma,
    \label{eqn7:criticalScalingNL}
\end{equation}
where the values are unchanged (we previously chose $\mathcal{K}=4$ and $\gamma=0.36$ as representative values during radiation domination). Because the compaction takes a maximum value $C_{max}$ at $C_{1,max}=4/3$, we now have a maximum mass PBH which can form for a given horizon mass,
\begin{equation}
    M_\mathrm{PBH,max} = \mathcal{K} M_\mathrm{H}\left( C_{max}-C_c \right)^{\gamma} \approx 2.1M_\mathrm{H},
\end{equation}
or, equivalently, the minimum horizon mass which can form a PBH of a certain mass,
\begin{equation}
    M_\mathrm{H,min}=\frac{M_\mathrm{PBH}}{\mathcal{K}\left( C_{max}-C_c \right)^{\gamma}} \approx 0.48 M_\mathrm{PBH}.
    \label{eqn7:minMh}
\end{equation}
Inverting equation \eqref{eqn7:criticalScalingNL} gives us $\nu$ as a function of the PBH mass,
\begin{equation}
    \nu(M_\mathrm{PBH}) = \frac{C_1(M_\mathrm{PBH})}{\sigma_0} = \frac{2}{3\sigma_0}\left(2-\sqrt{4-6C_c-6\left( \frac{M_\mathrm{PBH}}{\mathcal{K}M_\mathrm{H}} \right)^{1/\gamma}} \right),
\end{equation}
where we have kept only the solution for type I perturbations, and $\sigma_n$ is given by equation \eqref{eqn7:momentsP}. We will also require the derivative,
\begin{equation}
    \frac{\mathrm{d}\nu}{\mathrm{d}M_\mathrm{PBH}} = \frac{\sqrt{2}\left( \frac{M_\mathrm{PBH}}{\mathcal{K}M_\mathrm{H}}\right)^{\frac{1}{\gamma}-1}}{\gamma\mathcal{K}M_\mathrm{H}\sqrt{2-3C_c-3\left(\frac{M_\mathrm{PBH}}{\mathcal{K}M_\mathrm{M_H}}\right)^{\frac{1}{\gamma}}}}.
\end{equation}

Using the theory of peaks, the initial abundance of PBHs collapsing to form PBHs is given by,
\begin{equation}
    \beta(r) = \frac{4\pi r^3}{3} \int\limits_{\nu_{c}}^{\frac{4}{3\sigma_0}} \mathrm{d}\nu \mathcal{K}\left( \nu\sigma_0-\frac{3}{8}\left( \nu\sigma_0 \right)^2-C_c\right)^\gamma\frac{1}{3^{3/2}(2\pi)^2}\frac{\sigma_1^3}{\sigma_0^3}\exp\left( -\frac{\nu^2}{2} \right),
\end{equation}
We have here included the upper limit for PBH formation, corresponding to $C_\mathrm{max}=4/3$, although this typically has a minimal effect on the PBH abundance. 

Again, the total PBH abundance $f_\mathrm{PBH}$ and the mass function $\psi_\mathrm{PBH}$ can then be obtained following the same method as described in section \ref{sec7:PS};
\begin{equation}
    f_\mathrm{PBH}=\frac{\Omega_\mathrm{PBH}}{\Omega_\mathrm{CDM}} =\frac{1}{\Omega_\mathrm{CDM}} \int\limits_{M_\mathrm{min}}^{M_\mathrm{max}} \frac{\mathrm{d}M_\mathrm{H}}{M_\mathrm{H}}\left(\frac{M_\mathrm{eq}}{M_\mathrm{H}}\right)^{1/2}\beta(M_\mathrm{H}),
    \label{eqn7:omegaPBHNL}
\end{equation}
where equation \eqref{eqn7:horizonMass} relates the horizon mass $M_\mathrm{H}$ to the (smoothing) scale $r$. The integral runs over the entire range of horizon masses for which PBH formation occurs.

The PBH mass function is, likewise, calculated in the same fashion as previously,
\begin{multline}
    \psi(M_\mathrm{PBH})=M_\mathrm{PBH} \frac{\mathrm{d}f_\mathrm{PBH}}{\mathrm{d} M_\mathrm{PBH}}= \frac{M_\mathrm{PBH}}{\Omega_\mathrm{CDM}}\int\limits_{M_\mathrm{H,min}}^{M_\mathrm{max}} \frac{\mathrm{d}M_\mathrm{H}}{M_\mathrm{H}}\left(\frac{M_\mathrm{eq}}{M_\mathrm{H}}\right)^{1/2}\\
    \times\frac{4\pi r^3(M_\mathrm{H})}{3} \frac{\mathrm{d}\nu}{\mathrm{d}M_\mathrm{PBH}} \frac{M_\mathrm{PBH}}{M_\mathrm{H}}\frac{1}{3^{3/2}(2\pi)^2}\frac{\sigma_1^3}{\sigma_0^3}\exp\left( -\frac{\nu^2(M_\mathrm{PBH})}{2} \right).
\end{multline}
It is here important to note that we now integrate from the minimum horizon mass which forms a PBH of mass $M_\mathrm{PBH}$, given by equation \eqref{eqn7:minMh}. It is now important not to neglect this limit, as the peak height $\nu$ becomes imaginary for smaller values.



\section{Summary}
\label{sec7:summary}

We have discussed how cosmological perturbations collapse to form PBHs at horizon entry, with larger mass PBHs corresponding to larger scale perturbations formed at earlier times during cosmological inflation. Because PBHs form deep in the radiation dominated epoch, and are constrained to be less abundant than dark matter, they are very rare upon formation, making up only a small fraction of the energy content of the universe. The abundance of PBHs depends upon the amplitude of perturbations in the early universe, with larger amplitude perturbations resulting in a higher PBH abundance. This means that the PBH abundance can be calculated from the primordial power spectrum, providing a unique method to constrain the early universe via constraints on the PBH abundance. An accurate calculation of the PBH abundance also means that predictions can be made for other interesting factors, such as the mass function and initial clustering.

There are numerous ways in which perturbations can be parameterised, including the curvature perturbation, the density contrast, but in the context of PBH formation, it is typically most useful to consider perturbations described by the compaction. A perturbation is considered to collapse to form a PBH if the compaction peaks above a certain threshold value, $C_c$. The compaction function is conserved on super-horizon scales, and the statistics are relatively simple, even when non-linearities are accounted for. However, as perturbations re-enter the horizon, their evolution becomes significantly more complicated - although a linear transfer function is often used to describe the evolution on horizon and sub-horizon scales, the validity of which can be questioned~\cite{DeLuca:2023tun}. It is therefore preferable to calculate the PBH abundance in the super-horizon regime by comparing the initial amplitude of perturbations (long before horizon entry) to the critical value. However, depending on the power spectrum being considered, this is not always possible.

The most important parameter for determining the PBH abundance is the the threshold value for collapse. The critical value is sufficiently large that no PBHs are expected to form in the universe, if the power spectrum is not significantly larger on small scales than is visible in the CMB. The exact value of the threshold depends upon the profile shape of a given perturbation, but since perturbations are generally expected to be close to the Mexican-hat profile, it is safe to use the critical value corresponding to that profile, $C_c=0.5$.

Several methods for calculating the PBH abundance and mass function have been described, with the theory of peaks being the most common method. Extensions to the theory of peaks do exist, but these typically do not offer a significant improvement in the accuracy of such calculations, and can be far more complex. 

Throughout this chapter, we have assumed an initial Gaussian distribution for the curvature perturbation $\zeta$, but in section \ref{sec7:non-linearities} the effect of non-linearity between $\zeta$ and the density contrast is considered. This non-linearity results in the compaction following a non-Gaussian distribution function, which significantly increases the amplitude of the power spectrum required to form a given abundance of PBHs.

\bibliography{main}
\end{document}